\documentstyle[11pt]{article}
\begin{document}
\date{} %optional
%\begin{titlepage}
%%%%
\begin{flushright}
{MIT-CTP-3138}\\
{YITP-01-39}
\end{flushright}
%%%%
\vspace{2cm}
\begin{center}
%%%%
{\Large {\bf Comment on Kaluza-Klein Spectrum \\
 of Gauge Fields in the Bigravity Model}
%%%%%
\vskip0.8truein
{\large Motoi Tachibana}~\footnote{E-mail:{\tt motoi@lns.mit.edu}
~~JSPS Research Fellow\\
on leave from Yukawa Institute for Theoretical Physics,
Kyoto University, Kyoto 606-8502, Japan}\\}
\vskip0.4truein
\centerline{ {\it 
Center for Theoretical Physics,
Massachusetts Institute of Technology, Cambridge, MA 02139, USA.}}
\end{center}
%%%%%
\vskip0.8truein
%%%%%
%\baselineskip=0.5 truein plus 2pt minus 1pt
%%%% 
%\twocolumn[\maketitle\abstract{
\centerline{\bf Abstract}
\vskip0.3truein
We study behavior of bulk gauge field in the bigravity model 
in which two positive tension $AdS_4$ branes in $AdS_5$ bulk 
are included.  We solve the equations of motions for Kaluza-Klein 
modes and determine the mass spectrum. It is shown that unlike
the case of graviton, we find no ultralight Kaluza-Klein modes
in the spectrum.

\vskip0.13truein
\baselineskip=0.5 truein plus 2pt minus 1pt
%}]
\baselineskip=18pt
%%%%
%\baselineskip=0.5 truein plus 2pt minus 1pt
%%%%%
%\vskip 2cm
%\end{titlepage}
\addtolength{\parindent}{2pt}
\newpage
%%%%
Over the past few years, theories with extra spatial dimensions
have received much attention because they potentially
solve long standing problems such as the hierarchy problem,
as was originally suggested by Antoniadis,
Arkani-Hamed, Dimopoulos and Dvali\cite{add,aadd}.
Afterwards Randall and Sundrum found a new solution to the
hierarchy problem\cite{rs1}. In addition, they showed that gravity 
could be localized on a brane with infinite extra dimension\cite{rs2}.
Since then, various types of models including gravity have been studied.

Recently Kogan et al. proposed a new \lq \lq bigravity" 
model\cite{kogan}. The model
has only two positive tension $AdS_4$ branes in $AdS_5$ bulk and no 
negative tension branes. Owing to the absence of negative tension branes,
the model does not have the ghost fields which appeared 
in the previous \lq \lq bigravity"
scenario\cite{kmprs} and also in the quasi-localized gravity
model\cite{grs}. But interestingly in the model of \cite{kogan}
the bounce of the warp factor mimics the effect of a negative 
tension brane and thus gives rise to an
anomalously light graviton Kaluza-Klein mode. Moreover it is possible
in this model to circumvent 
the van Dam-Veltman-Zakharov no go theorem\cite{dv,z}
about the non-decoupling of the extra polarization states of the massive
graviton\cite{kmp2}. Similar discussions have recently appeared in 
\cite{kr}-\cite{sch}.

In the previous work\cite{motoi1}, we studied behavior of the zero
mode of the bulk gauge field in this bigravity model in more detail. 
In this paper, we try to understand the massive (Kaluza-Klein) 
modes in the model. Scalar and fermion fields in this model
have already been discussed extensively\cite{chacko,mouslopoulos}.
It has already  wellknown that 
bulk gauge fields in the
ordinary Randall-Sundrum model have quite a specific behavior.
The reason consists in the conformal
property of the gauge field action, {\it i.e.,} the 4-dimensional
kinetic term of the gauge field is not warped, which leads us to 
the fact that the 
zero mode of the bulk gauge field in the RS model is not 
localized on a brane.  It is flat in the extra dimension.
Many people have tried to resolve this 
issue\cite{dvali}-\cite{oda3},
but no one has not succeeded completely yet. In that meaning, it is
important to investigate various aspects of bulk gauge fields 
in different higher dimensional models including the bigravity model.

Following the work by Kogan et al.\cite{kogan}
we shall consider a five dimensional
anti-de Sitter spacetime ($AdS_5$) with a warp factor
%%%%
\begin{equation}
ds^2 = \Omega^2(w)(\bar{g}_{\mu \nu}(x)dx^{\mu} dx^{\nu} + dw^2),
\label{metric}
\end{equation}
%%%%
where $\mu, \nu = 0,1,2,3.$ 
The metric $\bar{g}_{\mu \nu}(x)$
denotes an $AdS_4$ background. The warp factor $\Omega(w)$ in the
conformal coordinate $w$ is given by
%%%%
\begin{equation}
\Omega(w)= \frac{1}{\cosh(kz_0)}\frac{1}{\cos\tilde{k}(|w|-\theta)},
\label{warp}
\end{equation}
%%%%
where $\tilde{k} \equiv k/\cosh(kz_0)$ and
$\tan(\tilde{k}\theta/2)=\tanh(kz_0/2)$, while $\tanh(kz_0)
\equiv kV_1/|\Lambda|$. $k$ is the curvature of $AdS_5$ defined
through $k \equiv \sqrt{\frac{-\Lambda}{24M^3}}$. $\Lambda$ is
the five dimensional cosmological constant
which is negative. $V_1$ and $V_2$
are tensions of the 3-branes at the orbifold fixed points,
$w=0$ and $w=w_L$, respectively.
\newpage
We are interested in behavior of the bulk gauge field in this background
metric. Let us start with the following five dimensional action;
%%%%
\begin{equation}
S_{GF}=-\frac{1}{4}\int d^4x\int_{-w_L}^{w_L}dw
\sqrt{-G}G^{MN}G^{RS}F_{MR}F_{NS},
\label{action}
\end{equation}
%%%%
where $M, N, R, S= 0,1,2,3,w$ and
$F_{MN}=\partial_M A_N -\partial_N A_M$. $A_M(x^{\mu}, w)$
is the bulk U(1) gauge field. The extension to non-Abelian gauge
field is straightforward. $G_{MN}$ is the five dimensional metric
defined through eq.(\ref{metric}).
The equations of motions are given by
%%%%
\begin{equation}
\partial_M(\sqrt{-G}G^{MN}G^{RS}F_{NR})=0.
\label{eom}
\end{equation}
%%%%
We solve the equations with the gauge condition $A_w=0$ and
as usual, we expand the field $A_{\mu}(x^{\mu}, w)$ into zero mode
and Kaluza-Klein modes:
%%%%
\begin{equation}
A_{\mu}(x, w)=\sum_{n=0}^{\infty}a^{(n)}_{\mu}(x)\rho^{(n)}(w).
\label{kk}
\end{equation}
%%%%
Then the equation of motion for the bulk gauge field
reduces to\cite{motoi1}
%%%%
\begin{equation}
\partial_w (\Omega(w)\partial_w \rho^{(n)})+m_n^2\Omega(w)\rho^{(n)}=0.
\label{eom3}
\end{equation}
%%%%
Here $m_n$ denote the mass eigenvalues.

The zero mode of the gauge field ($m_n=0$) is of the form
%%%%
\begin{equation}
\rho^{(0)}(w)=\frac{a}{\tilde{k}}
\cosh(kz_0)\sin\tilde{k}(w-\theta) + b,
\label{zmsol}
\end{equation}
%%%%
where $a$ and $b$ are integration constants. 
We discussed some properties of the zero mode in the
previous paper\cite{motoi1} so that we don't mention
it any more.

On the other hand, the KK modes satisfy the following equations;
%%%%
\begin{eqnarray}
\frac{d^2 \rho^{(n)}}{dw^2}+\tilde{k}\tan(\tilde{k}(w-\theta))
\frac{d\rho^{(n)}}{dw}+m_n^2 \rho^{(n)}&=&0 \qquad
{\rm for} \quad w>0, \nonumber \\
\frac{d^2 \rho^{(n)}}{dw^2}+\tilde{k}\tan(\tilde{k}(w+\theta))
\frac{d\rho^{(n)}}{dw}+m_n^2 \rho^{(n)}&=&0 \qquad 
{\rm for} \quad w<0.
\label{kkeq2}
\end{eqnarray}
%%%%
%Here we define $\rho(w) \equiv 
%\cos^{\frac{1}{2}}(\tilde{k}w)\tilde{\rho}(w)$
%and plug it into the above equations, giving
%%%%
%\begin{equation}
%\frac{d^2 \tilde{\rho}^{(n)}(x)}{dx^2}+
%\Bigl(c_1 + \frac{c_2}{\cos^2 x}\Bigr)\tilde{\rho}^{(n)}(x)=0,
%\label{kkeq3}
%\end{equation}
%%%%
%where 
%%%%%
%\begin{eqnarray}
%c_1&=& \biggl(\frac{m_n}{\tilde{k}}\biggr)^2+\frac{1}{4}, \nonumber \\ 
%c_2&=&-\frac{3}{4}, \nonumber \\
%x &\equiv& \tilde{k}(w \mp \theta).
%\label{kkeq4}
%\end{eqnarray}
%%%%
The general solution of Eqs.(\ref{kkeq2}) is given 
in terms of hypergeometric functions\cite{motoi1};
%%%%
\begin{eqnarray}
\rho^{(n)}(w)&=&c_1 F\Bigl(\alpha_n, \beta_n, \frac{1}{2};
\sin^2 (\tilde{k}(|w|-\theta))\Bigr) \nonumber \\
& & \qquad \qquad +c_2 |\sin\tilde{k}(|w|-\theta)|
F\Bigl(\alpha_n +\frac{1}{2}, \beta_n +\frac{1}{2}, \frac{3}{2};
\sin^2 (\tilde{k}(|w|-\theta))\Bigr).
\label{hyper}
\end{eqnarray}
%%%%
Here $c_1$ and $c_2$ are integration constants and the mode-dependent
parameters $\alpha_n$ and $\beta_n$ are given by
%%%%
\begin{eqnarray}
\alpha_n &=& -\frac{1}{4}
+\frac{1}{2}\sqrt{\biggr(\frac{m_n}{\tilde{k}}\biggr)^2+\frac{1}{4}}, 
\nonumber \\
\beta_n &=& -\frac{1}{4}
-\frac{1}{2}\sqrt{\biggl(\frac{m_n}{\tilde{k}}\biggr)^2+\frac{1}{4}}.
\label{hyper2}
\end{eqnarray}
%%%%
Similar expressions have been obtained in the cases of 
graviton\cite{kogan} and fermions\cite{mouslopoulos}. 
Below, we shall consider the case of symmetric 
configuration, {\it i.e.,} $w_L = 2\theta$ for simplicity.
In this case, we have only to consider
even and odd functions with respect to the minimum of the
warp factor. In the case of the odd functions, we have $c_1=0$
while the even functions, we have $c_2=0$. Indeed
%%%%
\begin{eqnarray}
\rho_{even}^{(n)}(w)&=&c_1 F\Bigl(\alpha_n, \beta_n, \frac{1}{2};
\sin^2 (\tilde{k}(|w|-\theta))\Bigr), \nonumber \\
\rho_{odd}^{(n)}(w)&=& c_2 |\sin\tilde{k}(|w|-\theta)|
F\Bigl(\alpha_n +\frac{1}{2}, \beta_n +\frac{1}{2}, \frac{3}{2};
\sin^2 (\tilde{k}(|w|-\theta))\Bigr).
\label{hyper3}
\end{eqnarray}
%%%%
In order to obtain the mass spectrum, we have to impose the
boundary conditions. In this case they should be given as follows;
%%%%
\begin{equation}
\rho_{even}^{(n)'}(0)=\rho_{even}^{(n)'}(2\theta)=\rho_{odd}^{(n)'}(0)
=\rho_{odd}^{(n)'}(2\theta)=0
\label{boundary}
\end{equation} 
%%%% 
where the prime denotes the derivative against $z$. 

We evaluate 
Eqs.(\ref{hyper3}) under the boundary conditions (\ref{boundary}).
%at large $kz_0$ limit (or equivalently $\tilde{k}\theta \rightarrow
%\pi/2$ limit). 
For the odd modes, we find
%%%%
\begin{equation}
-\frac{1}{3}\Bigl(\frac{m}{k}\Bigr)^2\sinh^2(kz_0)
F\Bigl(\alpha+\frac{3}{2},\beta+\frac{3}{2},\frac{5}{2};\tanh^2(kz_0)
\Bigr)
+F\Bigl(\alpha+\frac{1}{2},\beta+\frac{1}{2},\frac{3}{2};\tanh^2(kz_0)
\Bigr)=0.
\label{spectrumodd}
\end{equation}
%%%%
Here we neglected the suffix $n$ and used the relation $\sin(\tilde{k}
\theta) = \tanh(kz_0)$.
%%%%
%\begin{equation}
%m_n^{(odd)} = \tilde{k}\sqrt{e^{2kz_0}+6-\frac{2}{1-4e^{-2kz_0}}}
%\sim k
%\label{spectrumodd}
%\end{equation}
%%%%
%since $\tilde{k} \sim ke^{-kz_0}$ at large $kz_0$. 

On the other hand, for the even modes
%%%%
\begin{equation}
F\Bigl(\alpha+1,\beta+1,\frac{3}{2};\tanh^2(kz_0)\Bigr)=0.
\label{spectrumeven}
\end{equation}
%%%%
Now let us evaluate the eqs.(\ref{spectrumodd}) and
(\ref{spectrumeven}) at large $kz_0$ limit to obtain
the mass spectrum. As the result, we can analytically
find the following expressions for the mass of the KK mode.
For odd modes
%%%%
\begin{equation}
m_n = 2\sqrt{(n+1)(n+\frac{3}{2})}ke^{-kz_0}
\label{oddmass}
\end{equation}
%%%%%
where we have used an useful formula of the hypergeometric function
%%%%
\begin{equation}
F(\alpha, \beta, \gamma; 1) = \frac{\Gamma(\gamma)
\Gamma(\gamma-\alpha-\beta)}
{\Gamma(\gamma-\alpha)\Gamma(\gamma-\beta)}
\label{formula}
\end{equation}
%%%%
$\Gamma(x)$ is the Euler's gamma function .

On the other hand, the KK mass spectrum for even modes 
at $kz_0 >> 1$ is obtained as  
%%%%
\begin{equation}
m_n = 2\sqrt{(n+1)(n+\frac{1}{2})}ke^{-kz_0}
\label{evenmass}
\end{equation}
%%%%%
Note here that this behavior of the mass spectrum is quite 
different from the case of graviton\cite{kogan}. 
We find no ultralight Kaluza-Klein modes in the spectrum
of the gauge field. The reason seems to be that in the 
present case, the bulk gauge field corresponds
to $\nu = \frac{1}{2}$ in Ref.\cite{mouslopoulos}.

What happens in the asymmetric case where $w_L \neq 2\theta$?
In this case, we also expect that the qualitative property
still holds, i.e., the KK mass spectrum becomes the order of $ke^{-kz_0}$.
But we have one more parameter in this case, which is the distance
between the two $AdS_4$ branes, $w_L$. 
%If $w_L << 2\theta$, 

To conclude,
in this paper we studied the behavior of the bulk gauge fields
in the bigravity model. We solved the equations of motions for
the Kaluza-Klein gauge fields and obtained the mass spectrum at
large distance ($kz_0$) limit. As the result, we had no ultralight
Kaluza-Klein modes in the mass spectrum unlike the case of
graviton.

%Next we shall consider the effective field theory 
%in terms of the zero mode on the $AdS_4$
%brane. For that purpose we plug eq.(\ref{zmsol}) into the
%original action (\ref{action}). The result is 
%\cite{oda}
%%%%
%\begin{eqnarray}
%S_{GF}^{(0)} &=& -\frac{1}{4}\int dx^4\int_{-w_L}^{w_L}dw
%\sqrt{-G}G^{MN}G^{RS}F_{MR}^{(0)}F_{NS}^{(0)} \nonumber \\
%&=& -\frac{1}{4}\int dx^4
%\sqrt{-\bar{g}}\bar{g}^{\mu \nu}\bar{g}^{\lambda \sigma}
%f_{\mu \lambda}^{(0)}f_{\nu \sigma}^{(0)}
%\int_{-w_L}^{w_L}dw\Omega(w)(\rho^{(0)}(w))^2 \nonumber \\
%& &-\frac{1}{4}\int dx^4 \sqrt{-\bar{g}}\bar{g}^{\mu \nu}
%a_{\mu}^{(0)}a_{\nu}^{(0)}\int_{-w_L}^{w_L}dw
%2\Omega(w)(\partial_w \rho^{(0)}(w))^2.
%\label{zmaction}
%\end{eqnarray}
%%%%

\vspace{1cm}
%\newpage
\begin{center}
{\large{\bf Acknowledgments}}
\end{center}
We would like to thank Andreas Karch and Stavros Mouslopoulos
for useful conversations and comments. 
This work was supported in part by a Grant-in-Aid for Scientific 
Research from Ministry of Education, Science, Sports and Culture 
of Japan (No.~3666).

\end{document}